\documentclass{article}
\usepackage{amssymb}
\usepackage{amsmath,bm}
\usepackage{array}
\usepackage{caption,graphicx}
\usepackage{xcolor}

\newcommand{\be}{\begin{equation}}
\newcommand{\ee}{\end{equation}}

\def\un{{\rm 1\mkern-4mu I}}

\begin{document}

\title{\bf Superselection of the weak hypercharge and the algebra of the Standard Model }

\author{{Ivan TODOROV}
	\small 
	\\
	\\ Institut des Hautes \'Etudes Scientifiques, 35 route de Chartres, 
	\\ F-91440 Bures-sur-Yvette, France
	\\
	\\
	and
	\\
	\\ Institute for Nuclear Research and Nuclear Energy\footnote{Permanent address.}, \\ Bulgarian Academy of Sciences 
	\\ Tsarigradsko Chaussee 72, BG-1784 Sofia, Bulgaria
	\\ ivbortodorov@gmail.com}

\maketitle 
	

\bigskip

	\begin{abstract}

Restricting the $\mathbb{Z}_2$-graded tensor product of Clifford algebras $C\ell_4\hat{\otimes}C\ell_6 $ to the particle subspace allows a natural definition of the Higgs field $\Phi$, the scalar part of Quillen's superconnection, as an element of 
$C\ell_4^1$. We emphasize the role of the exactly conserved weak hypercharge Y, promoted here to a superselection rule for both observables and gauge transformations. This yields a change of the definition of the particle subspace adopted in recent work  with Michel Dubois-Violette \cite{DT20}; here we exclude the zero eigensubspace of Y consisting of the sterile (anti)neutrinos which are allowed to mix. One thus modifies the Lie superalgebra generated by the Higgs field. Equating the normalizations of $\Phi$ in the lepton and the quark subalgebras we obtain a relation between the masses of the W boson and the Higgs that fits the experimental values within one percent accuracy.


	 
	\end{abstract}
	
	\newpage
	
	\tableofcontents
	
	
	
	\newpage
\section{Introduction}\label{sec1}

The attempts to understand "the algebra of the Standard Model (SM) of particle physics" started with the Grand Unified Theories (GUT) (thus interpreted in the illuminating review \cite{BH}), was followed by a vigorous pursuit by Connes and collaborators
of the noncommutative geometry approach to the SM (reviewed in \cite{CC, S}). The present work belongs to a more recent development, initiated by Dubois-Violette \cite{DV} and continued in \cite{TD, TDV, DT, T}, that exploits the theory of euclidean Jordan algebras (see also \cite{BF, B}).  We modify the superconnection associated with the Clifford algebra $ C\ell_{10}$ considered in \cite{DT20}. A fresh look on the subject is offered with a special role assigned to the exactly conserved elctroweak hypercharge Y, which commutes with both observables and gauge transformations. But first, some motivation.


The spinor representation of the grand unified theory $Spin(10)$
	\begin{equation} 
		\underline{32}= \underline{16}_L +\underline{16}_R \label{1}
		\end{equation}
fits perfectly one generation of fundamental (anti)fermions of the Standard Model. Its other representations, however, have no satisfactory physical interpretation. For instance, the 45-dimensional adjoint representation  involves leptoquarks (on top of the expected eight gluons and four electroweak gauge bosons) and predicts unobserved proton decay. The Clifford algebra $C\ell_{10}$, whose derivations span the Lie algebra $so(10)$, 
has, on the other hand, a single irreducible representation (IR) which coincides with  (\ref{1}). 
The chirality operator $\chi$ can be identified with the Coxeter element $\chi= \omega_{9,1} (=i\omega_{10}$) of the real form $C\ell(9,1)$ of $C\ell_{10}=C\ell(10, {\mathbb C})$. It has the property to commute with the even part $C\ell_{10}^0$ of 
$C\ell_{10}$, which contains $so(10)$, and anticommutes with its odd part. The Higgs field intertwines between left and right chiral fermions and will be associated with a suitable projection of the odd part of the Clifford algebra.

The complexification of the underlying algebra allows to display the duality between observables and symmetry transformations. The important obsevables, both external (like energy momentum) and internal (charge, hypercharge) are conserved. Conservation laws are related to symmetries by Noether theorem. Continuous internal symmetries are generated by antihermitian elements of Lie algebras of compact groups. Observables, on the other hand, correspond to hermitian (selfadjoint) operators. In the non-exceptional case one should deal with a complex (associative) \textit{field algebra} (to borrow the term of Haag \cite{H}) that contains both observables and symmetry generators. Then the algebraic statement of Noether's theorem will result in identifying the conserved observables with symmetry generators multiplied by the imaginary unit $i=\sqrt{-1}$  - see \cite{K13,F19} as well as the discussion in \cite{B20}.

In order to formulate the quark-lepton symmetry it would be convenient to view $C\ell_{10}$ as a $Z_2$-graded tensor product of Clifford algebras generated by Fermi oscillators $a_\alpha^{(*)} (= a_\alpha$ or $a_\alpha^*, \alpha =1, 2)$ and $b_j^{(*)}, j=1, 2, 3$, respectively:
\begin{equation}
\label{Cl4timesCl6}
C\ell_{10}=C\ell_4\hat{\otimes}C\ell_6, \, a_\alpha^{(*)}\in C\ell_4, b_j^{(*)}\in C\ell_6, [a_\alpha^{(*)}, b_j^{(*)}]_+=0,
\end{equation}
$\alpha$ playing the role of a flavour (weak isospin) and $j$ of a colour index. (The above tensor product  has been 
earlier introduced by Furey \cite{F} within the division algebra approach to the SM.) A distinguished element of an oscillator algebra is the number operator. The difference of normalized number operators,
\be 
\label{a*a-b*b}
\frac{1}{2} Y=\frac{1}{3}\sum_{j=1}^3b_j^*b_j-\frac{1}{2}\sum_{\alpha=1}^2a^*_\alpha a_\alpha, \, \,
\ee
is the exactly conserved (half) weak hypercharge.  To insure the quark-lepton (colour-flavour) symmetry we shall promote it to a superselection rule\footnote{Superselection rules were introduced by G.C. Wick, A.S. Wightman and E.P. Wigner \cite{WWW}; superselection sectors in algebraic quantum field theory were studied by Haag and collaborators (see Sect, IV.1 of \cite{H}). For a pedagogical review and further references - see \cite{G}.}: all observables and gauge Lie algebra generators are assumed to be invariant under the following (global) $U(1)_Y$ phase transformation of $a_\alpha^{(*)}$ and $b_j^{(*)}$:
\begin{equation}
\label{U(1)Y}
a_\alpha\rightarrow e^{\frac{i}{2}\varphi} a_\alpha \, (a^*_\alpha \rightarrow e^{-\frac{i}{2}\varphi} a_\alpha^*), \, b_j\rightarrow e^{-\frac{i}{3}\varphi}b_j, \, \alpha=1, 2;  j=1, 2, 3, \varphi \in \mathbb{R}.
\end{equation}
This requirement yields the gauge Lie subalgebra $\mathfrak{g}= u(2)\oplus u(3)\subset so(10)$ that does not involve leptoquark gauge fields\footnote{For a different approach to unification without leptoquarks - see \cite{KS}.} and leads to a non-simple internal observable algebra (see Sect. 2 and Appendix A). We note that $Y$ annihilates the rank two Jordan subalgebra $J_{s\nu}$ of sterile (anti)neutrinos $\nu_R, \bar{\nu}_L$. The maximal subalgebra of $\mathfrak{g}$ that annihilates $J_{s\nu}$ is the gauge Lie algebra of the SM:
\be 
\label{GSM}
\mathfrak{g}_{SM} = su(3)_c \oplus su(2)_L \oplus u(1)_Y \subset \mathfrak{g}=u(2)\oplus u(3)\subset so(10), \, \mathfrak{g}_{SM} J_{s\nu} = 0.
\ee
The extra $u(1)$ term in $\mathfrak{g}$, not present in the gauge Lie algebra of the SM, can be identified with the difference $B-L$ of baryon and lepton numbers (or, equivalently, with twice the third component of the right chiral isospin $2I_3^R$):
\be
\label{B-L}
B-L=\frac{1}{3}\sum_{j=1}^3[b_j^*, b_j] = Y-2I_3^R, \, 2I_3^R=\frac{1}{2}\sum_{\alpha=1}^2 [a_\alpha, a_\alpha^*] = a_1 a_1^*-a_2^* a_2.
\ee
We have $2I^R_3\nu_R=\nu_R=-(B-L)\nu_R \Rightarrow Y\nu_R=0$.

 The algebra of $U(1)_Y$-invariant elements contains besides the obvious products $a_\alpha^*a_\beta, b_j^*b_k$ also the isotropic elements
\be
\label{Omega}
 \Omega=a_1a_2b_1b_2b_3, \, \Omega^*=b_3^*b_2^*b_1^*a_2^*a_1^*, \, \, \Omega^2 = 0 = (\Omega^*)^2,
 \ee
whose products are idempotents corresponding to the sterile neutrino states:
\be 
\label{AntiNu}
\Omega \Omega^*=\nu_R, \, \Omega^* \Omega=\bar{\nu}_L.
\ee
We can only distinguish particles from antiparticles with $Y\neq 0$. The sterile neutrino and antineutrino have not been
observed and we expect them to "oscillate" - and mix (see the pioneer paper \cite{P}).  By definition, observables span a Jordan algebra\footnote{The finite dimensional euclidean Jordan algebras are classified in \cite{JvNW} (for a concise review see Sect. 2 of \cite{T}). Their role in the present context has been emphasized in \cite{DV}.} of hermitian operators that commute with all superselection charges (in our case with $Y$). The $U(1)_Y$-invariant Jordan subalgebra $J$ of $C\ell_{10}$ splits into three pieces: the particle, $J_\mathcal{P}$, and the antiparticle $J_{\bar{\mathcal{P}}}$ parts (with $Y\neq 0$ each) and the rank two subalgebra $J_{s\nu}$ of sterile (anti)neutrinos.

The forces of the SM have two ingredients, the gauge fields and the Higgs boson, liken to the Beauty and the Beast of the fairy tale in a popular account \cite{M}. The superconnection that includes the Higgs field is an attempt to transform the Beast into Beauty as well. An effective superconnection has been used by physicists (Ne'eman, Fairly) since 1979 - see, especially, \cite{T-MN}, before the mathematical concept was coined by Quillen \cite{Q, MQ}. A critical review of the involuted history of this notion and its physical implications is given in Sect. IV of \cite{T-M} (see also Sect. I of \cite{T-M20}). (One should also mention the neat exposition of \cite{R} - in the context of the Weinberg-Salam model with two Higgs doublets.) The state space of the SM is $Z_2$ graded - into left and right chiral fermions - and the Higgs field intertwines between them. It should thus belong to the odd part of the underlying Clifford algebra which anticommutes with chirality $\chi$ (satisfying $\chi^2=1$). The exterior differentials entering the connection form $D=d+A=dx^\mu(\partial_\mu+A_\mu)$ anticommute. As noticed by Thierry-Mieg \cite{T-M} if we replace $D$ by $\chi D=D\chi$ it will also anticommute with the Higgs field which belongs to the odd part of the Clifford algebra (Sect. 3). This change does not alter the classical curvature $D^2$ as 
$\chi^2=1$.

We begin in Sect. 2 by recalling the relation between the Fermi oscillator realization of even (euclidean) Clifford algebras and isometric complex structures in $C\ell_{2\ell}$, \cite{D}. There we also introduce the basic projectors $\pi_\alpha^{(')}=\pi_\alpha$ or $\pi_\alpha^{'}, \alpha = 1, 2$ and $p_j^{(')}, j=1, 2,3$. A complete set of commuting observables is given by five traceless linear combinations of these projectors. Their $2^5$ 5-element products give a complete set of primitive idempotents describing the states of fundamental (anti)fermions in one generation.   The decomposition of the Jordan algebra of $U(1)_Y$ invariant observables into simple components, displayed in Sect. 2, is discussed in more detail in the Appendix. 

Sect. 3 starts by reproducing a result of \cite{DT20}: projecting on the particle subspace kills the possible colour components $b_j^{(*)}$ of the Higgs field, thus guaranteeing that gluons remain massless. The exclusion of the sterile neutrino from the projector  $\mathcal{P}$ on the particle subspace transforms the Fermi oscillators $a_\alpha^{(*)}$ into the odd generators of the simple Lie superalgebra\footnote{Let us warn the reader that, unlike the popular Lie superalgebras whose representations feature unobserved superpartners of known bosons and fermions, the even and odd parts of Ne'eman-Fairly $sl(2|1)$ representations correspond to the familiar right and left chiral leptons and quarks.} $sl(2|1)$, a unexpected new result. Previously, the same Lie superalgebra (called $su(2|1)$) has been proposed on the basis of the observation that only the supertrace of Y vanishes in the space of leptons (see the review in Sect. I of \cite{T-M20}). In Sect. 4 we first display the existence of a massless photon in the unitary gauge (an alternative derivation of this result within the superconnection approach has been given in \cite{R}). We also reproduce the result of \cite{DT20} on the Weinberg angle and the ensuing ratio between the masses of the W and Z bosons. A surprizing new result of Sect. 4 is the relation $m_H=2cos\theta_W m_W$ between the Higgs and W masses and the \textit{theoretical value} of the Weinberg angle, verified within one percent accuracy for the observed values of the masses and the value $4cos^2\theta_W=\frac{5}{2}$. After a brief survey of the chiral fermionic Lagrangian and the condition for absence of a "scalar anomaly" in Sect. 5   
we summarize and discuss the results in Sect. 6.

\bigskip

\section{Fock space realization of $C\ell_4\hat{\otimes}C\ell_6$} \label{sec2}
\setcounter{equation}{0}
The complexification $E_c$ of a $2\ell$-dimensional real euclidean space $E$ with a (positive) scalar product $( , )$ admits s family of $\ell$-dimensional isotropic subspaces, in one-to-one correspondence with skew-symmetric orthogonal transformations 
$J: (x,Jy)=-(Jx, y), J^2=-1$. Each such $J$ defines a linear \textit{complex structure} - see \cite{D}. For each splitting of an orthonormal basis $e_1, ..., e_{2\ell}$ into two complementary sets $I$ and $I^{'}$ of $\ell$ elements,
we can define a $J$ such that $Je_j = e_j^{'}, \, Je_j{'}=-e_j, e_j\in I, e_j^{'}\in I^{'}$. Then the two conjugate sets of $\ell$ elements
$$
n_j=\frac{1}{2}(e_j+iJe_j), \, \bar{n}_j= \frac{1}{2}(e_j-iJe_j), e_j\in I, \, n_j, \bar{n}_j\in E_c,
$$
satisfy 
$$
 (n_j, n_k)=0=(\bar{n}_j, \bar{n}_k), (\bar{n}_j, n_k)=(\bar{n}_k, n_j)=\delta_{jk}, \, \, \,   Jn_j=-in_j, J\bar{n}j=i\bar{n}_j.
$$
If $\gamma: E\rightarrow C\ell(E)$ is the map of $E$ to the generators of the $2^{\ell}$ dimensional spinor representation of the Clifford algebra, such that 
\begin{equation}
\label{gamma(x)}
[\gamma(x), \gamma(y)]_+:= \gamma(x)\gamma(y)+\gamma(y)\gamma(x)=2(x, y)\textbf{1},
\end{equation}
extended by linearity to $E_c$, then setting $\gamma(n_i)=f_i, \gamma(\bar{n}_i)=f_i^*$ the $f_i^{(*)}(=f_i$\, or \, $f_i^*)$ satisfy the canonical anticommutation relations (CAR):
$$
[f_i, f_j]_+=0=[f_i^*, f_j^*]_+, \, [f_i, f_j^*]_+=\delta_{ij}.
$$
The complexified space $E_c$ has a natural notion of complex conjugation that preserves $E$ and can hence be equipped with a sesquilinear Hilbert space scalar product such that
\begin{equation}
\label{HilbSc}
<x, y>=(\bar{x}, y), \, <x, x> \, >0 \, \, for \,  x\neq 0, \, <x, y>= \overline{<y, x>}. 
\end{equation}
 As a result the complexified Clifford algebra $C\ell(E_c)=C\ell(E)\otimes\mathbb{C}$ admits a hermitian conjugation $A\rightarrow A^*$, an antilinear antihomomorphism such that
\begin{equation}
\label{H-star}
\gamma(x)^*=\gamma(\bar{x}),  x\in E_c; \, \, (AB)^*=B^*A^*, \, A, B\in C\ell(E_c).
\end{equation} 

As stated in  the introduction, we regard $C\ell_{10}$ as the $Z_2$-graded tensor product (\ref{Cl4timesCl6}) where the Fermi oscillators obey the CAR:
\begin{equation}
\label{CAR}
[a_\alpha, a_\beta]_+=0(=[a_\alpha^*, a_\beta^*]_+), [b_j, b_k]_+=0, \, [a_\alpha, a_\beta^*]_+ =\delta_{\alpha \beta}, 
[b_j, b_k^*]_+=\delta_{jk}
\end{equation}  
($\alpha, \beta = 1,2; \, j, k=1, 2, 3$). The Lie subalgebra of $so(10)$, invariant under the superselection rule (\ref{U(1)Y}), is a $u(1)$ extension of the Lie algebra of the SM:
\begin{eqnarray}
\label{u1u3}
\mathfrak{g}=u(2)\oplus u(3), \, \, u(2)= Span\{[a_\alpha^*, a_\beta], \alpha, \beta=1,2\}, \nonumber \\ 
u(3)=Span\{[b_j^*, b_k], j, k=1,2,3\}. 
\end{eqnarray}
In particular, the weak (left) isospin components $I_\sigma(=I^L_\sigma)$ are given by:
\be 
\label{isospin}
I_+=a_1^*a_2, I_-=a_2^*a_1, 2I_3=[I_+,I_-]=a_1^*a_1-a_2^*a_2 \,.
\ee
A maximal set of commuting observables is generated by five pairs of mutually orthogonal projectors:
\begin{equation}
\label{projectors}
\pi_\alpha =a_\alpha a_\alpha^*, \pi_\alpha^{'}=a_\alpha^*a_\alpha = 1-\pi_\alpha; \, p_j=b_jb_j^*, p_j^{'}=b_j^*b_j=1-p_j; 
\pi_\alpha\pi_\alpha^{'}=0=p_jp_j^{'},
\end{equation}
$\alpha=1, 2; j=1, 2, 3$.  The $2^5$ products $\pi_1^{(')}\pi_2^{(')}p_1^{(')}p_2^{(')}p_3^{(')}$  with different distribution of primes provide a complete set of (rank one) \textit{primitive idempotents} which include all (pure) (anti)fermion states.  The projections on non-zero left and right isospin $P_1$ and $P_1^{'}$ are mutually orthogonal:
\be 
\label{P1}
P_1 = [I_+, I_-]_+ = (2I_3)^2 = \pi_1\pi_2^{'} + \pi_1^{'}\pi_2, \, P_1^{'}=(2I_3^R)^2= \pi_1 \pi_2 + \pi_1^{'}\pi_2^{'} 
\ee  
($P_1+P_1^{'}=1, P_1P_1^{'}=0$). 
The electric charge operator,
\be 
\label{Q}
Q= \frac{1}{2}Y + I_3 = \frac{1}{3}\sum_{j=1}^3b_j^*b_j - a_2^*a_2,
\ee
commutes with $a_1^{(*)}$ which will single out the neutral component of the Higgs field.  We note that while there is no coherent superposition of states of different charges (just as there is none of different $Y$'s), there are charge carrying (non-abelian) gauge fields, like $W_\mu^+I_+ + W_\mu^-I_-$, while, according to the $U(1)_Y$ superselection rule, there are none non-commuting with $Y$.

The left and right chiral fermion subalgebras $J_\mathcal{P}^L$ and
$J_\mathcal{P}^R$ of $J_\mathcal{P}$ have a rather different structure: $J_\mathcal{P}^L$ is the sum of two simple Jordan subalgebras of rank 6 and 2, while $J_\mathcal{P}^R$ splits into three simple pieces of rank 3, 3, 1 (see the Appendix):
\be 
\label{JPLR}
J_\mathcal{P}=J_\mathcal{P}^L\oplus J_\mathcal{P}^R, \, J_\mathcal{P}^L=J_6^2\oplus J_2^2, J_\mathcal{P}^R=J_3^2\oplus
 J_3^2 \oplus \mathbb{R} e_R
\ee
where $J_r^2= \mathcal{H}_r(\mathbb{C})$ (we use the notation of \cite{T}, Sect. 2.2). The $u(2)$ Lie algebra spanned by $\Omega, \Omega^*$ and their products (\ref{AntiNu}) is the projection of the right chiral isospin. We will not discuss its possible role in neutrino physics in this paper. We just note that being associated with $J_{s\nu}$ it completes the observed duality between IRs of compact Lie algebras and simple components of the Jordan algebra of superselected observables. Recall that a Majorana mass term in $J_{s\nu}$ violates both $B-L$ and $2I_3^R$ but still preserves $Y$. 

The algebras $J_\mathcal{P}$ and $J_{\bar{\mathcal{P}}}$ are isomorphic mirror images of one another, their elements differing by the signs of $Y, Q, B-L$, so it suffices to consider $J_\mathcal{P}$.  

We proceed to list the primitive idempotents with their (internal space) fermion pure states interpretation. To begin with, there are two rank four $SU(3)$-invariant (colourless) projectors on leptons and antileptons:
\be 
\label{colourless}
\ell=p_1 p_2 p_3, \, \bar{\ell}=p_1^{'} p_2^{'} p_3^{'}, \, (B-L+1)\ell=0=(B-L-1)\bar{\ell}, \, tr\ell=4=tr\bar{\ell}.
\ee
The pure lepton states in $J_\mathcal{P}$ are identified by the eigenvalues of the pair $(Q, Y)$:
\be 
\label{Leptons}
\nu_L=\pi_1^{'}\pi_2\ell \, (0,-1), \, e_L=\pi_1\pi_2^{'}\ell \, (-1,  -1); \,  e_R=\pi_1^{'} \pi_2^{'} \ell \, (-1, -2).
\ee
The sterile (anti)neutrino have both $Y=0=Q$ and only differ by the chirality 
\be 
\label{chirality}
\chi:=[a_1, a_1^*][a_2, a_2^*] [b_1, b_1^*] [b_2, b_2^*] [b_3, b_3^*] \,  (=\omega_{9,1});
\ee
\be 
\label{nuRL}
\nu_R=\pi_1\pi_2\ell = \Omega\Omega^*, \, \, \bar{\nu}_L = \Omega^*\Omega, (\chi-1)\nu_R=0=(\chi+1)\bar{\nu}_L.
\ee
There are three more rank four projectors $q_j, j=1,2,3$ on the subspaces of quarks of colour $j$ and any flavour:
\be 
\label{U(b)}
q_j:= U(b_j, b_j^*) \bar{\ell}= b_j \bar{\ell}b_j^* = p_jp_k^{'} p_\ell^{'} \, \, (U(x,y)z:=xzy+yzx),
\ee
where $(j, k, \ell$) is a permutation of $(1, 2, 3)$.  The pure quark states are:
\be 
\label{quarks}
u^j_L= \pi_1^{'} \pi_2 q_j, \, d_L^j=\pi_1\pi_2^{'}q_j; \, \, u_R^j=\pi_1\pi_2 q_j, d_R^j=\pi_1^{'}\pi_2^{'}q_j.
\ee 
In fact, since $SU(3)_c$ is an exact gauge symmetry individual colour states are not observed. One should introduce instead 
gauge-invariant density matrices; the sum $\mathfrak{q}=q_1+q_2+q_3$ is also an idempotent (since the $q_j$ are mutually orthogonal) and is $SU(3)_c$ invariant. Thus one can use the density matrices (by definition of trace one) obtained from (\ref{quarks}) by replacing $q_j$ with $\frac{1}{3}\mathfrak{q}$. 

In view of (\ref{Leptons}), (\ref{quarks}) the 15 dimensional projector $\mathcal{P}$ on the particle subspace can be written as
the projector $\mathcal{P}_0$ used in \cite{DT20} minus $\nu_R$:
\be  
\label{P15}
\mathcal{P} = \mathcal{P}_0 - \Omega\Omega^*, \, \mathcal{P}_0 = \ell + \mathfrak{q}, \, \Omega\Omega^* = \ell \pi_1\pi_2(=\nu_R).
\ee
We shall see that this modification changes the Lie superalgebra generated by the Higgs superconnection in an interesting way.

\bigskip

\section{Particle subspace, Higgs field and associated Lie superalgebra}\label{sec3}
\setcounter{equation}{0}
A general problem in theories, whose configuration space is a product of a commutative algebra of (continuous) functions on space-time with a finite dimensional quantum algebra, is the problem of fermion doubling \cite{GIS} (still discussed over twenty years later, \cite{BS}). It was proposed in \cite{DT20}, as a remedy, to consider the algebra $\mathcal{P}_0 C\ell_{10}\mathcal{P}_0$ where $\mathcal{P}_0$ is the projector on the 16 dimensional particle subspace including the right handed sterile neutrino $\nu_R$ (see (\ref{P15})). Note that the 16 dimensional subspace of the fundamental representation \textbf{27} of the $E_6$ GUT is also commonly identified with the space of particles (see e.g. \cite{B}). As recalled in the introduction $\nu_R$ and its antiparticle $\bar{\nu}_L$ both belong to the zero eigenspace of $\mathfrak{g}_{SM}$ and are allowed to mix by the $U(1)_Y$ superselection rule (as they do in the popular theory involving a Majorana neutrino). We shall use instead the 15-dimensional projector $\mathcal{P}= \mathcal{P}_0 - \nu_R$ (\ref{P15}). This will lead to changing the projection of the flavour Lie superalgebra on the lepton subspace. 

We shall first display the projection of the factor $C\ell_6$ in (\ref{Cl4timesCl6}) which does not change when substituting 
$\mathcal{P}_0$ by $\mathcal{P}$. To begin with, as $\mathcal{P} \mathcal{P}_0 = \mathcal{P}$, the odd part of $C\ell_6$, killed by $\mathcal{P}_0$, is annihilated a fortiori by $\mathcal{P}$:
\be 
\label{PbP0} 
\mathcal{P}_0 b_j^{(*)} \mathcal{P}_0 = 0 \Rightarrow \mathcal{P} b_j^{(*)} \mathcal{P} = 0.
\ee
The generators of $su(3)$ change in a way that preserves their commutation relations (CRs). We proceed  to displaying 
$\mathcal{P}\frac{1}{2}[b_j^*, b_k] \mathcal{P}$. Let again $(j, k, l)$ be a permutation of $(1, 2, 3)$; then $\frac{1}{2}[b_j^*, b_k]=b^*_j b_k$ and, using(\ref{U(b)}), we find: 
\be 
\label{Pbj*k}
B_{jk}:=\mathcal{P}b_j^* b_k\mathcal{P}= \mathfrak{q} b_j^*b_k\mathfrak{q}  = q_kb_j^*b_k q_j=b_j^*b_k p_\ell^{'}.
\ee
The preservation of the CRs then follows from the relations:
\be 
\label{Pbjkl}
[B_{jk}, B_{kl}]=b_j^*b_\ell p_k^{'} = B_{j\ell}, \, \, \,  \mathcal{P}(p_j^{'}-p_k^{'})\mathcal{P}=(p_j^{'} -p_k^{'})p_\ell^{'} = q_k-q_j.
\ee

Novel things happen when projecting the first factor, $C\ell_4$ in (\ref{Cl4timesCl6}). The projection of the Fermi oscillators $a_\alpha^{(*)}$ is nontrivial. Indeed, the easily verifiable relations $a_\alpha \pi_\alpha=0, \pi_\alpha a_\alpha = a_\alpha; \, \pi_\alpha a_\alpha^*=0, a_\alpha^*\pi_\alpha=a_\alpha^*$  imply:
\begin{eqnarray}
\label{ProjA}
\mathcal{P}a_\alpha\mathcal{P}=\mathfrak{q} a_\alpha+\ell (1-\pi_1\pi_2)a_\alpha, \, \mathcal{P}a_\alpha^*\mathcal{P}=\mathfrak{q} a_\alpha^*+\ell a_\alpha^*  (1-\pi_1\pi_2) , \nonumber \\
\ell a_1^{(*)}\rightarrow \ell a_1^{(*)}\pi_2^{'},  \, \, \, \ell a_2^{(*)}\rightarrow \ell a_2^{(*)}\pi_1^{'}. \, \,
\end{eqnarray} 
It turns out that the resulting odd elements of $C\ell_4$ can be identified with the odd generators of the Lie superalgebra 
$s\ell(2|1)$ (also denoted as $su(2|1)$ - see \cite{T-M20}). Indeed, using the conventions of (Sect. 3.1 of) \cite{GQS} and setting
\be 
\label{sl21odd} 
F_+=-a_2\pi_1^{'}, F_-=-a_1\pi_2^{'}, \, \bar{F}_+=a_1^*\pi_2^{'}, \bar{F}_- = a_2^*\pi_1^{'}, \, 2Z=-\pi_1^{'}-\pi_2^{'}(=\ell Y,
\ee
we recover the super CRs of $s\ell(2|1)$ :
$$
[F_+, F_-]_+=0=[\bar{F}_+, \bar{F}_-]_+, \, [F_\pm, \bar{F}_\pm]_+=I_\pm, \, [F_\pm, \bar{F}_\mp]_+=Z\mp I_3; 
$$
$$
[I_+, I_-] = 2I_3, [2I_3, F_\pm]=\pm F_\pm, [2I_3, \bar{F}_\pm]=\pm \bar{F}_\pm, \, [Z, I_\pm]=0=[Z, I_3];
$$
$$
[I_\pm, F_\pm]=0=[I_\pm, \bar{F}_\pm], \, [I_\pm, F_\mp]=-F_\pm, \, \, [I_\pm, \bar{F}_\mp]=\bar{F}_\pm,  \,
$$
\be 
\label{ACR}
[2Z, F_\pm]=F_\pm, \, \, [2Z, \bar{F}_\pm] = -\bar{F}_\pm \, \, (F_\pm^* = -\bar{F}_\mp). \,  
\ee
On the other hand, the projection $\mathfrak{q} a_\alpha^{(*)}$ of $a_\alpha^{(*)}$ satisfies the unmodified CARs (\ref{CAR}). As a result it is simpler to display the associated lepton and quark representation spaces separately (omitting the projectors $q$ and $\ell$).

The 3-dimensional lepton subspace is \textit{atypical degenerate representation} of $s\ell(2|1)$ (see Sect. 3.2 of \cite{GQS})
with highest weight state $\pi_1^{'}\pi_2^{'}$ (annihilated by $\bar{F}_\pm$). The lepton state vectors $|Y, 2I_3>, Y=2Z=-\pi_1^{'}-\pi_2^{'}$ are given by:
\begin{eqnarray}
\label{3RepLep}
|e_R>=\ell a_1^*a_2^*=:|-2, 0>, \bar{F}_\pm|-2, 0>=0, \, F_\pm|-2, 0>=:\pm|-1,\pm 1>, \nonumber \\
  |-1, \pm 1>= |\nu_L>/|e_L>; \,  I_\pm|-1, \pm 1>=0, \, I_\pm|-1, \mp 1> =|-1, \pm 1>. \, \,
\end{eqnarray}
We note that only the projectors $\pi_{1,2}^{'}=-Z\pm I_3$ are defined in the Lie superalgebra $s\ell(2|1)$. The complementary projectors $\pi_\alpha:=1-\pi_\alpha^{'}$ do not have the same trace:
\be 
\label{Tr(pi)}
tr 1=3, \, \pi_{1,2}^{'}=I_3\mp Z \Rightarrow tr\pi_\alpha^{'}=2, \, \, \, tr \pi_\alpha=tr(1-\pi_\alpha^{'})=1.
\ee
The three 4-dimensional mutually orthogonal projectors $q_j, j=1, 2, 3,$ give rise to isomorphic $u(2)\rtimes CAR_2$ modules. For ease of notation we shall omit the subscript $j$ on $q, d_R, u_L, ...$. We note that the two spinor doublets $(a_1^*, a_2^*)$ and $(-a_2, -a_1)$ transform under commutation with $u(2)$ in the semidirect product $u(2)\rtimes CAR_2$,  in the same way as $\bar{F}_\pm$ and $F_\pm$ above.  The quark space $\mathcal{H}_q$ can be obtained by acting on the highest weight ket vector $qa_1^*a_2^*$ (annihilated by the left action of the raising operators $a_\alpha^*$) with polynomials of the lowering operators $a_\alpha$. It is four dimensional with $(Y, 2I_3)$ basis: 
\begin{eqnarray}
\label{QuarkStates}
|d_R>=|-\frac{2}{3}, 0>:=qa_1^*a_2^*, \, |u_L>=|\frac{1}{3}, 1>:= -a_2|-\frac{2}{3}, 0> =qa_1^*\pi_2; \, |d_L>= \nonumber \\
|\frac{1}{3},-1>:=a_1|-\frac{2}{3}, 0> = \pi_1a_2^*, \, |u_R>=|\frac{4}{3}, 0>:= a_2a_1|-\frac{2}{3}, 0>=q\pi_1\pi_2 .\, \, \,
\end{eqnarray}
Here we have used the general formula for the hypercharge $\mathcal{P}Y\mathcal{P}=\frac{4}{3}\mathfrak{q} - \pi_1^{'} - \pi_2^{'} $.

\textit{Remarks} 1. The representation spaces of coloured quarks and leptons appear as minimal left ideals in the enveloping algebra of the respective Lie superalgebra (cf. \cite{Ab}). Individual ket vectors are elements $X$ of the algebra, related to the corresponding idempotents of the (euclidean) Jordan subalgebra of $C\ell_4\otimes C\ell_6^0$ by $X\rightarrow XX^*$; for instance,
$$
|\nu_L>= \ell a_1^*\pi_2\rightarrow \nu_L=|\nu_L><\nu_L|= \ell \pi_1^{'}\pi_2, \, 
$$
\be 
\label{KetBra}
|d_R>=q a_1^* a_2^*\rightarrow d_R= |d_R><d_R|=q\pi_1^{'}\pi_2^{'}, \, |u_L>=qa_1^*\pi_2\rightarrow u_L=q\pi_1^{'}\pi_2. 
\ee
2. The trace of the hypercharge $Y$ takes equal values in the left and right chiral subspaces (-2 for leptons, 2/3 for a quark of fixed colour). Only their difference, the \textit{supertrace} vanishes for a given IR of the Lie superalgebra. The sum of $trY$ for all left (or right) chiral particle IRs (leptons and three coloured quarks)  does vanish, reflecting the cancellation of anomalies between quarks and leptons.

3. The $s\ell(2|1)$ realization of the antileptons is somewhat tricky and  we shall spell it out (although we won't use it later). To begin with the antiparticle projector that excludes the sterile antineutrino reads:
\be 
\label{antileptonProjection}
 \bar{\mathcal{P}}=\bar{\mathcal{P}}_0-\bar{\nu}_L = \bar{\ell}(1-\pi_1^{'}\pi_2^{'})+\bar{\mathfrak{q}}, \, \bar{\ell}=p_1^{'}p_2^{'}p_3^{'}, \, \bar{\mathfrak{q}}=\sum\bar{q}_j, \, \bar{q}_j.=p_j^{'}p_kp_\ell,
\ee
$(j, k, \ell)\in Perm(1, 2, 3)$. We find $\bar{\mathcal{P}}a_\alpha^{(*)}\bar{\mathcal{P}} = \bar{\ell}a_\alpha^{(*)}\pi_{\bar{\alpha}} +\bar{\mathfrak{q}}a_\alpha^{(*)}, \, \alpha=1, 2, \bar{\alpha}=3-\alpha$. Thus we arrive at $F_+=-a_2\pi_1,F_-=-a_1\pi_2, \bar{F}_+=a_1^*\pi_2, \bar{F}_-=a_2^*\pi_1$. Finally we apply the outer automorphism (correcting on the way Eq. (3.6) of \cite{GQS})
\be 
\label{outer} 
I_\pm, I_3; Z; F_\pm, \bar{F}_\pm, \, \rightarrow \, I_\pm, I_3; -Z; \pm\bar{F}_\pm, \pm F_\pm ,
\ee
ending up with the antilepton $s\ell(2|1)$ generators (labelled by a superscript $^a$):
$$
I_\pm^a= \bar{\ell} I_\pm \, I_3^a=\bar{\ell} (\pi_1 - \pi_2); \, 2Z^a=\bar{\ell}(\pi_1+\pi_2)=\bar{\ell}Y; 
$$
\be 
\label{AntiGenerators}
F_+^a=\bar{\ell} a_1^*\pi_2, F_-^a=-\bar{\ell}a_2^*\pi_1, \bar{F}_+^a = -\bar{\ell}a_2\pi_1, \bar{F}_-^a= \bar{\ell}a_1\pi_2 .
\ee
Note that the conjugation properties are not preserved by the outer automorphism (\ref{outer}): while $F_\pm^*=-\bar{F}_\mp$, we have $(F_\pm^a)^*=\bar{F}_\mp^a$ for the antileptons.
  
We observe that the the ket vectors (\ref{3RepLep}) (\ref{QuarkStates}) of left chiral fermions belong to the odd subspace $C\ell_4^1\otimes C\ell_6^0$ of the Clifford algebra, while the right chiral fermion kets belong to its even subalgebra $C\ell_4^0\otimes C\ell_6^0$. The (antihermitian) Higgs component of the superconnection
\be 
\label{Higgs}
\Phi(x)=\ell(\phi_1 \bar{F}_++\bar{\phi}_1 F_- +\phi_2\bar{F}_- +\bar{\phi}_2 F_+) + \rho \mathfrak{q} \sum_\alpha(\phi_\alpha a_\alpha^*-\bar{\phi}_\alpha a_\alpha) \, 
 \ee
$(\phi_\alpha=\phi_\alpha(x))$ is odd and intertwines the left and right subspaces. The normalization factor $\rho$ in front of $\mathfrak{q}$ will be fixed later.
 Following the suggestion of \cite{T-M} we include the chirality $\chi$ in the definition of superconnection:  
\be 
\label{ConnectPhi}
\mathbb{D} = \chi D +\Phi, \, D=d+A=dx^\mu(\partial_\mu +A_\mu), \, iA_\mu=W_\mu+B_\mu+G_\mu;
\ee
here $W=W^+I_++W^-I_-+W^3I_3, \, B$ is proportional to $Y, G(\in su(3))$ is the gluon field spanned by $B_{jk}$ (\ref{Pbj*k}) and the differences $q_j-q_k$. Since $[\chi, D]=0=[\chi, \Phi]_+$ the canonical supercurvature $\mathbb{F}_0=i\mathbb{D}^2$ involves, as it should,  the commutator (rather than the anticommutator) of $A$ and $\Phi$; recalling that $\chi^2=1$ we find:
\be 
\label{Curvature}
-i\mathbb{F}_0:=\mathbb{D}^2 = D^2+\chi[D, \Phi] +\Phi^2, \, [D, \Phi] = dx^\mu(\partial_\mu\Phi+[A_\mu, \Phi])
\ee
($iD^2$ is spanned by hermitian matrix valued fields: $iF_{\mu\nu}=F_{\mu\nu}^aT_a, T_a^*=T_a$).  
\be 
\label{HiggsCurv}
\Phi^2= \ell(\phi_1\bar{\phi}_2 I_++\bar{\phi}_1\phi_2 I_- -\phi_1\bar{\phi}_1\pi_2^{'}-\phi_2\bar{\phi}_2 \pi_1^{'}) -\rho^2\mathfrak{q}\phi \bar{\phi}, \, \, \phi\bar{\phi}= \phi_1\bar{\phi}_1 + \phi_2\bar{\phi}_2     
\ee 
(the Higgs curvature is $i\Phi^2$). As further discussed in \cite{T-M} the \textit{Bianchi identity}
\begin{equation}
\label{Bianchi}
\mathbb{D}\mathbb{F}_0 = (\chi(d+A)+\Phi)\mathbb{F}_0- \mathbb{F}_0(\chi A+\Phi) =0
\end{equation}
for the supercurvature $\mathbb{F}_0=i\mathbb{D}^2$, an expression of the associativity relation
\be 
\label{associativity}
\mathbb{D}\mathbb{D}^2 = \mathbb{D}^2 \mathbb{D},
\ee
is equivalent to the (super) \textit{Jacobi identity} for our Lie superalgebra. We note that (\ref{Bianchi}) amounts to three equations, one for each power of $\Phi$:
$$
[D, D^2] = 0 \Leftrightarrow [A, dA] + dA^2 =0  \, \,
$$
$$
[\Phi, D^2] +[\chi D,\chi (d\Phi+[A, \Phi])]=0 \Leftrightarrow [\Phi, D^2]+D[D, \Phi] +[D, \Phi]D=0, 
$$ 
\be 
\label{3Bianchi}
[\chi D, \Phi^2] + [\Phi, \chi[D, \Phi]] = \chi(d\Phi^2 + [A, \Phi^2] - [\Phi, D\Phi]_+) = 0.
\ee
The relation $[A, \Phi^2]=[\Phi, [A, \Phi]]_+$ which enters the last equation (\ref{3Bianchi}) follows from the super Jacobi identity for two odd generators $F_1, F_2$ and one even, $A$,
\be 
\label{SuperJacobi}
[A, [F_1, F_2]_+] +[F_1, [F_2, A]]_+ = [F_2, [A, F_1]]_+,
\ee
by setting $F_1=F_2=\Phi$.

 Happily, the Bianchi identity still holds if we add to $i\Phi^2$ a constant term:
\be 
\label{Fm}
\mathbb{F}_0 \rightarrow \mathbb{F}= \mathbb{F}_0 + i\hat{m}^2, \, \hat{m}^2= m^2(\ell(1-\pi_1\pi_2) + \rho^2 \mathfrak{q}).
\ee
Only for $m^2>0$ shall we have a non trivial minimum  of the classical bosnic action and the gauge bosons will acquire a non zero mass.  

\bigskip

\section{Bosonic Lagrangian; mass relations}
\setcounter{equation}{0}
The action density corresponding to the curvature $\mathbb{F}$ (\ref{Fm}) is proportional to the product of $\mathbb{F}$ with its Hodge dual of its hermitian conjugate $*\mathbb{F}^*$ (for a textbook exposition, see Sect. 7.2 of \cite{H17}). We shall write the action density and the  corresponding bosonic Lagrangian in the form:
$$
\mathcal{L}(x)dV = -Tr(\mathbb{F} *\mathbb{F}^*), \, dV = dx^0dx^1dx^2dx^3(:=dx^0\wedge dx^1\wedge dx^2\wedge dx^3),
$$
$$
\mathbb{F}=i(D^2+\chi[D,\Phi] +\Phi^2+\hat{m}^2) \Rightarrow \mathbb{F}^*=i(D^2+\chi[D,\Phi] -\Phi^2 - \hat{m}^2);
$$ 
\be 
\label{Lagrangian}
\mathcal{L}(x)=Tr\{\frac{1}{2}F_{\mu\nu}F^{\mu\nu} - (\partial_\mu\Phi+[A_\mu, \Phi])(\partial^\mu\Phi+[A^\mu, \Phi])\} - V(\Phi) .
\ee
We proceed to explain and write down in more detail each term in (\ref{Lagrangian}). $iF_{\mu \nu}$ is the sum of three gauge field strengths corresponding to $W, B, G$ (\ref{ConnectPhi}). The trace $Tr$ is normalized in a way to have the standard expression for the gluon field strength: 
$$
\frac{1}{2}Tr(G_{\mu \nu}G^{\mu \nu})=-\frac{1}{4} G_{\mu \nu}^a G^{\mu \nu}_a, \, iG_{\mu \nu} = G_{\mu \nu}^aT_a, \, TrT_aT_b=\frac{1}{2}\delta_{a b}.
$$  
This yields a non trivial relation between $Tr$ and the \textit{Jordan trace} $tr$, normalized to take the value 1 for one dimensional projectors (primitive idempotents). Writing $G$ (omitting the tensor indices) in the form (cf. (\ref{Pbj*k}))
\be 
\label{GluonBasis}
G=\sum_{j\neq k} G^{jk}B_{jk} +G^3T_3+G^8T_8, \, B_{jk}=b^*_jb_k p^{'}_\ell, 2T_3 =q_1-q_2, 2\sqrt{3}T_8=q_1+q_2-2q_3
\ee
($(j, k, \ell)\in Perm(1, 2, 3)$) we shall have
\be 
\label{tr=4Tr} 
trG^2 = 4\sum_{j\neq k} G^{jk}G^{kj} + 2 ((G^3)^2 + (G^8)^2) = 4 TrG^2.
\end{equation}
The Higgs potential $V(\Phi)$ is given by
\be 
\label{HiggsPotential}
V(\Phi)= Tr(m^2(\ell(1-\pi_1\pi_2)+\rho^2\mathfrak{q})+\Phi^2)^2 -\frac{1}{4} m^4 = \frac{1}{2}(1+6\rho^4)(\phi\bar{\phi}-m^2)^2.
\ee
In deriving (\ref{HiggsPotential}) we have used $\pi_\alpha+\pi_\alpha^{'}=1$ and $[I_+, I_-]_+=\pi_1^{'}\pi_2+\pi_1\pi_2^{'}$ to find:
$$
\phi_1\bar{\phi}_1\pi_2^{'}+\phi_2\bar{\phi}_2\pi^{'}_1= \phi\bar{\phi}\pi_1^{'}\pi_2^{'} +\phi_1\bar{\phi}_1\pi_1\pi_2^{'}+
\phi_2\bar{\phi}_2\pi_1^{'}\pi_2, \, Tr\Phi^4 = \frac{1}{2}(1+6\rho^4)(\phi\bar{\phi})^2.
$$
 The subtraction of $\frac{1}{4} m^4$ ensures the vanishing of the potential at its minimum (needed to have a finite action at the corresponding constant value $\Phi_0$ of $\Phi$). 
 
\textit{Remark} The standard notation $\mu^2\phi^\dagger \phi -\lambda (\phi^\dagger \phi)^2$ for the contribution of the Higgs potential to the Lagrangian\footnote{See \textit{•Mathematical formulation of the standard model} in Wikipedia (of October 16, 2020).} of the SM corresponds in (\ref{HiggsPotential}) to $\mu^2=(1+6\rho^4)m^2, \, \lambda = \frac{1}{2}(1+6\rho^4)$. We define however the square of the vacuum expectation value $v^2$ of the Higgs field as the minimum in $\phi\bar{\phi}$ of $V(\Phi)$  thus obtaining $<\phi\bar{\phi}>=v^2=\frac{\mu^2}{2\lambda}=m^2$ that is half the accepted standard value.
  
We shall use the \textit{unitary gauge} in which only the neutral component of the Higgs field - which commutes with the electric charge $Q=\frac{1}{3}(p_1^{'}+p_2^{'}+p_3^{'}) -\pi_2^{'}$ (\ref{Q}) - survives. The CRs $[Q, a_2]=a_2$ and
\be 
\label{CR_Q} 
[Q, F_+]= F_+, [Q, \bar{F}_-]=-\bar{F}_-, \, [Q, F_-]=0=[Q, \bar{F}_+] = [Q, a_1^{(*)}],
\ee
imply $\phi_2(x)=0$ in the unitary gauge while $\phi_1(x)=:\phi_0(x)$ is real and $\phi_0=m$ minimizes the potential:
\be 
\label{Phi0}
\Phi_0(x) := (\ell (F_- + \bar{F}_+) + \rho \mathfrak{q}(a_1^*-a_1))\phi_0(x), \, \phi_0(x) (= \bar{\phi}_0(x))= m +H(x).
\ee
The first approximation to the gauge bosons' mass term is obtained by replacing the Higgs field in the square of the commutator $[iA_\mu, \Phi]$ by its minimizing operator value - with  $H(x)=0$ in (\ref{Phi0}) (or, more generally, setting $\phi_0\bar{\phi}_0=m^2$). The gluon field (\ref{GluonBasis})
commutes with $a_\alpha^{(*)}$ and hence remains massless, in  accord with the fact that the $SU(3)_c$ gauge symmetry is unbroken. Thus only the electroweak gauge field $A^{ew}$ contributes to the commutator $[iA, \Phi]$:
\be 
\label{WB}
[A_\nu, \Phi]=[A^{ew}_\nu, \Phi], \, iA_\nu^{ew} = W_\nu^+I_++W^-_\nu I_-  +W_\nu^3 I_3+\frac{N}{2}B_\nu Y;
\ee
here the constant N is chosen to make the trace norms of $2I_3$ and $NY$ equal:
\be 
\label{Nsquare}  
tr\mathcal{P}(2I_3)^2=2+2\times 3 =8, tr\mathcal{P}(NY)^2= N^2(2+\frac{2}{3} +4+\frac{16}{3}+\frac{4}{3})=\frac{40}{3}N^2.
\ee
It follows that N, to be identified with the tangent of the Weinberg angle, satisfies
\be 
\label{tangentWeinberg}
N^2 = \frac{Tr (2I_3)^2}{Tr Y^2}(=\frac{tr(2I_3)^2}{tr Y^2})=: tg^2\theta_W= \frac{3}{5}.
\ee
Using the CRs (\ref{ACR}) and the relation $Y=\frac{2}{3}\sum_{j=1}^3p_j^{'} -\pi_1^{'}-\pi_2^{'}$ (\ref{a*a-b*b}) we get
$$
[W_\nu^+I_+ + W_\nu^-I_-,, \Phi]= W^+_\nu(\phi_2(\ell\bar{F}_+ +\rho\mathfrak{q}a_1^*) +\bar{\phi}_1(\rho\mathfrak{q}a_2-\ell F_+)) 
$$
$$
+ W^-_\nu (\phi_1(\ell\bar{F}_- +\rho\mathfrak{q}a_2^*) + \bar{\phi}_2(\rho\mathfrak{q}a_1-\ell  F_-));
$$
$$
[W^3_\nu 2I_3 + NB_\nu Y,  \Phi]= (W^3_\nu-NB_\nu)(\phi_1(\rho\mathfrak{q}a_1^*+\ell\bar{F}_+) + \bar{\phi}_1(\rho\mathfrak{q}a_1-\ell F_-))
$$
$$
-(W^3_\nu +NB_\nu)(\phi_2(\rho\mathfrak{q}a_2^*+\ell\bar{F}_-) +\bar{\phi}_2(\rho\mathfrak{q}a_2-\ell F_+)),
$$
The corresponding squares and their traces have the form (omitting the summed up 4-vector index):
$$
Tr[W^+I_+ + W^-I_-, \Phi]^2 = W^+W^-Tr(\phi_1\bar{\phi}_1\pi_1^{'}+\phi_2\bar{\phi}_2\pi_2^{'}+ \rho^2\mathfrak{q}\phi\bar{\phi})
$$
$$
=\frac{1+6\rho^2}{4}(W^+W^-+W^-W^+)\phi\bar{\phi} \, \, (4Tr\pi_\alpha^{'}=tr\pi_\alpha^{'}=2, \, \phi\bar{\phi}= 
\sum_{\alpha=1}^2 \phi_\alpha\bar{\phi}_\alpha),
$$
\be 
\label{TrWPki}
Tr[W^3I_3+\frac{N}{2}BY, \Phi]^2=\frac{1+6\rho^2}{8}((W^3-NB)^2\phi_1\bar{\phi}_1 +(W^3++NB)^2\phi_2\bar{\phi}_2).
\ee
In the unitary gauge, for $\Phi=\Phi_0$, the quadratic form in $W^3, B$ becomes degenerate (corresponding to zero photon mass)  
and for $\phi_0\bar{\phi}_0=m^2$ we find
\be
\label{WZmass}
 Tr[iA, \Phi_0]^2 = \frac{1+6\rho^2}{4}m^2( \, W^+W^-+W^-W^+ + \frac{1}{2}(W^3-NB)^2 \, ).
 \ee
It follows that only one linear combination of the neutral vector fields, 
\be 
\label{Zboson} 
Z_\nu = cW^3_\nu -sB_\nu, \, \frac{s}{c} = tg\theta_W = N(=\sqrt{\frac{3}{5}}), c^2+s^2=1,   
 \ee
acquires mass $m_Z$, satisfying  
\be 
\label{massZ}
m_Z^2 = \frac{m_W^2}{c^2} = (1+N^2) m_W^2.
\ee
The orthogonal linear combination remains massless and will be identified with the photon field:
\be 
\label{photon}
\Gamma_\nu:= s W^3_\nu + c B_\nu (\Rightarrow Z^2+\Gamma^2=W_3^2+B^2, m_\gamma = 0).
\ee
\textit{Remark} Although we follow the standard terminology and speak of unitary \textit{gauge} it should be emphasized that the choice $\Phi=\Phi_0$ (\ref{Phi0}) has a physical consequence: the vanishing of the photon mass. More generally, that follows from the vanishing of the product $\phi_1\phi_2$, enforced  in our earlier work  by adding an extra term in the superpotential (in  Eq. (4.2) of \cite{DT20}).

The relations among gauge boson masses are independent of the normalization constant $\rho$.  The ratio $m_H^2/m_W^2$ (for $m_H$ the Higgs mass), however, does depend on $\rho^2$. Indeed, equating it to the ratio of the coefficients to $\phi\bar{\phi}$ and $W^+W^-+W^-W^+$ in the Taylor expansion of the Higgs potential (\ref{HiggsPotential}) and in (\ref{WZmass}), respectively, we find:
\be 
\label{HiggsMass}  
m_H^2 = 4\frac{6\rho^4+1}{6\rho^2+1}m_W^2.
\ee
We shall fix $\rho^2$ by demanding that the leptonic input to $Tr\Phi^2$ equals to the contribution of a single coloured quark
(as it does if we do not project out $\nu_R$):
$$
-Tr(\ell\Phi^2)=Tr(\phi_1\bar{\phi}_1\pi_1\pi_2^{'}+\phi_2\bar{\phi}_2 \pi_1^{'}\pi_2+ \phi\bar{\phi}\pi_1^{'}\pi_2^{'})=\frac{1}{2}\phi\bar{\phi} (=\frac{1}{2}(\phi_1\bar{\phi}_1+\phi_2\bar{\phi}_2)),
$$
\be 
\label{rhoSquare}
-Tr(\rho^2 q^j\Phi^2)=\rho^2 Tr(q^j\phi\bar{\phi})=\rho^2\phi\bar{\phi} \, \Rightarrow \rho^2 = \frac{1}{2}.
\ee
This choice of $\rho^2$ yields:
\be 
\label{cosWeinmerg}
\frac{6\rho^4+1}{6\rho^2+1}=\frac{5}{8}=cos^2\theta_W, \, m_H=2cos\theta_W m_W = \sqrt{\frac{5}{2}} m_W.
\ee
 The last relation (for $2cos\theta_W=\sqrt{\frac{5}{2}}$) is verified within one percent error. Previous calculations in the superconnection approach (see, e.g. \cite{R}) yield the much higher value $m_H=2m_W$ that is some 35 GeV/c above the mark.
An intermediate result (closer to the experimental value) is claimed in \cite{HLN}.

\textit{Remark} The equality of the ratio $5/8$ to the \textit{theoretical value} of $cos^2\theta_W$ may be fortuitous: while $N^2=tg^2\theta_W$ equals the ratio of squares of coupling constants and is therefore running with the energy, $\rho^2$  equals the ratio of normalizations of the same quantity $Tr\Phi^2$ in two spaces and needs not run, thus further emphasizing the significance of the last relation (\ref{cosWeinmerg}). 

\bigskip

\section{Fermionic Lagrangian; anomaly cancellation}
\setcounter{equation}{0}

Having the Yang-Mills connection $D$ (\ref{ConnectPhi}) and the Higgs field $\Phi$ it is straightforward to write down the fermionic part of the Lagrangian. We proceed in some detail in order to fix our conventions.

\smallskip

We are using spacelike metric $(\eta_{\mu\nu} = {\rm diag} (-,+,+,+))$ and Dirac matrices satisfying $[\gamma_{\mu} , \gamma_{\nu}]_+ = 2 \, \eta_{\mu\nu}$. As we shall work with chiral fermions, we choose a $\gamma_5$ diagonal basis in which
$$
\gamma_5 (= i \gamma_0 \gamma_1 \gamma_2 \gamma_3 = \gamma_1 \gamma_2 \gamma_3 \beta) = - \sigma_3 \otimes \un_2 = \begin{pmatrix} -\un &0 \\ 0 &\un \end{pmatrix}, \beta \gamma^{\mu} = i \begin{pmatrix} - \widetilde\sigma^{\mu} &0 \\ 0 &\sigma^{\mu} \end{pmatrix},
$$
\be
\label{eq5-1}
\widetilde\sigma^0 = \sigma_0 = -\sigma^0 = - \widetilde\sigma_0 = 1_2, \quad \sigma^j = \sigma_j = \widetilde\sigma_j , \quad j=1,2,3,
\ee
$\sigma_j$ being the Pauli matrices (cf. Appendix I of \cite{T-M20}); here $\beta = i \gamma^0 (=\beta^*)$ defines a $U(2,2)$ invariant hermitean form, so that $\gamma_{\mu}^* \beta = - \beta \gamma_{\mu}$ where the star stands for hermitian conjugation. The conditions (\ref{eq5-1}) still leave a $U(1)$ freedom $\gamma^{\mu} \to S(\varphi) \gamma^{\mu} S(\varphi)^*$, $S(\varphi) = \exp \left( \frac i2 \, \varphi \gamma_5 \right)$, $\varphi \in {\mathbb R}$, in the choice of $\gamma^{\mu}$. The basis $\beta = \sigma_1 \otimes \un_2$, $\gamma_j = -\sigma_2 \otimes \sigma_j$ corresponds to charge conjugation matrix $C = \beta \gamma_2$ (defined to obey $^t\gamma_{\mu} C = - C \gamma_{\mu}$, for $^t\gamma_{\mu}$ transposed to $\gamma_{\mu}$). The two-by-two sigma matrices $\sigma_{\mu} (=\sigma_{\mu}^{A \dot B})$, $\widetilde\sigma_{\mu} = (\widetilde\sigma_{\mu \dot A B})$, $A,B,\dot A , \dot B = 1,2$ are chosen to satisfy
\be
\label{eq5-2}
\sigma_{\mu} \, \widetilde\sigma_{\nu} + \sigma_{\nu} \, \widetilde\sigma_{\mu} = 2 \, \eta_{\mu\nu} \un_L \, , \ \widetilde\sigma_{\mu} \, \sigma_{\nu} + \widetilde\sigma_{\nu} \, \sigma_{\mu} = 2 \, \eta_{\mu\nu} \un_R \, , \ \un_L = \left(\delta_B^A\right), \ \un_R = \left(\delta_{\dot A}^{\dot B}\right).
\ee
The fermionic part of the Lagrangian for the first generation of leptons and quarks reads:
\begin{eqnarray}
\label{eq5-3}
&&{\mathcal L}_F = - \widetilde\psi (\gamma^{\mu} D_{\mu} + \Phi) \psi = i \left(\overline L \, \widetilde\sigma^{\mu} D_{\mu} L - \overline R \, \sigma^{\mu} D_{\mu} R \right) - \nonumber \\
&&-i \left( \overline L_e \langle e_L \, \vert \overline\phi_{\ell} \vert \, e_R \rangle R_e + \overline L_d \langle d_L \, \vert \overline\phi_q \vert \, d_R \rangle R_d + \overline L_u \langle u_L \, \vert \phi_q \vert \, u_R \rangle R_u \right) + \nonumber \\
&&+ i \left( \overline R_e \langle e_R \, \vert \phi_{\ell} \vert \, e_L \rangle L_e + \overline R_d \langle d_R \, \vert \phi_q \vert \, d_L \rangle L_d + \overline R_u \langle u_R \, \vert \overline\phi_q \vert \, u_L \rangle L_u \right),
\end{eqnarray}
where we have set
$$
\psi = \begin{pmatrix} L_f^A \\ R_{f \dot B} \end{pmatrix} , \quad f=e,d,u \, , \quad \phi_{\ell} = \ell \, \overline F_+ \, \phi_1 (x) \quad \overline\phi_{\ell} = - \ell \, F_- \, \overline \phi_1 (x) \, ,
$$
\be
\label{eq5-4}
\phi_q = {\mathfrak q} \, a_1^* \, \phi_1 (x) \, , \quad \overline\phi_q = {\mathfrak q} \, a_1 \, \overline \phi_1 (x) \, ,
\ee
and the bras and kets define the idempotents
\begin{eqnarray}
\label{eq5-5}
&&\vert e_R \rangle \langle e_R \vert = \ell \, \pi'_1 \pi'_2 \, , \quad \vert e_L \rangle \langle e_L \vert = \ell \, \pi_1 \pi'_2 \, ; \nonumber \\
&&\vert d_R \rangle \langle d_R \vert = {\mathfrak q} \, \pi'_1 \pi'_2 \, , \quad \vert d_L \rangle \langle d_L \vert = {\mathfrak q} \, \pi_1 \pi'_2 \, , \quad \vert u_R \rangle \langle u_R \vert = {\mathfrak q} \, \pi_1 \pi_2 \, , \nonumber \\
&&\vert u_L \rangle \langle u_L \vert = {\mathfrak q} \, \pi'_1 \pi_2 \quad \left( \langle e_R \vert \, \overline F_+ \vert \, e_L \rangle = \ell \, \pi_1 \pi'_2 \, , \ {\rm tr} \, \ell \, \pi_1 \pi'_2 = 1 \right)
\end{eqnarray}
(see (\ref{3RepLep}), (\ref{QuarkStates}) for a possible choice). The full Lagrangian for the three generations of quarks (and leptons) should also involve the CKM quark mixing matrix \cite{PDG} (and, perhaps the Pontecorvo-Maki-Nakagawa-Sataka matrix?).

\smallskip

The standard treatment of the axial vector anomaly cancellation also applies to our case (cf. Appendix I to \cite{T-M20}). We proceed to consider a pair of oppositely oriented triangle graphs with one vector and two Higgs lines (Fig.~1)

$$
\includegraphics[width=9cm]{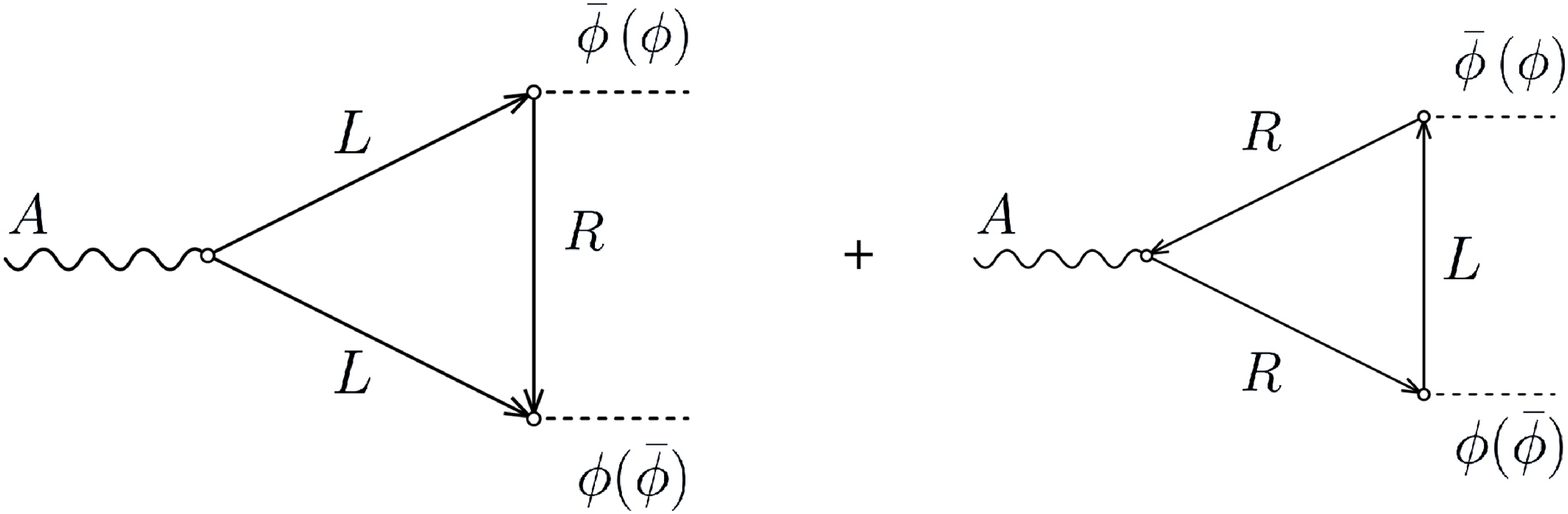}
$$
\centerline{{\bf Fig. 1.} \quad Scalar anomaly cancellation}

 \centerline{(The labels in parentheses correspond to a $u$-quark loop)}

\bigskip

\noindent that may involve a potential chiral ``scalar anomaly''. The fact that $\bar{\phi}$ and $\phi$ carry opposite values of $Y$ ($Y=1$ and $Y=-1$ - cf. (\ref{U(1)Y})) ensures hypercharge conservation in each vertex. A straightforward analysis shows that whenever {\it the supertrace} of the charge carried by the gauge field $A$ {\it vanishes} the divergence in the amplitudes of the two graphs of Fig.~1 cancels out. (The conclusion of Sect.~2 of \cite{T-M20} holds true for any Lie superalgebra, not just for $s\ell (2 |1)$, and hence applies to our treatment of the quark sector.) For instance, the total trace of the hypercharge of up and down quarks in the left and right sector coincides,
\be
\label{eq5-6}
{\rm tr} \, Y \vert_{\rm Left} = 6 \times \frac13 = 2 = {\rm tr} \, Y \vert_{\rm Right} = 3 \times \frac43 - 3 \times \frac23 =2 \, ,
\ee
so that their difference, the supertrace, vanishes. This is also true for the electric charge in both the quark and the lepton sectors but it fails for the leptonic hypercharge unless we assume a Higgs triggered transition between $\nu_L$ and the sterile neutrino -- a possible manifestation of neutrino oscillation whose study goes beyond the scope of the present paper.

\bigskip

\section{Summary and discussion}
\setcounter{equation}{0}

Our starting point was the observation that the unique (faithful) IR of the Clifford algebra $C\ell_{10}=C\ell_4\hat{\otimes} 
C\ell_6$ accommodates precisely the 32 fermion and antifermion states of a single generation of fundamental particles (including the hypothetical sterile (anti)neutrino needed to explain the observed neutrino oscillations). The complexified Clifford algebras $C\ell_{2\ell}=C\ell(2\ell, \mathbb{C})$ can be viewed as generated by Fermi oscillators  $a_\alpha^{(*)}\in C\ell_4,
\alpha=1, 2; \, b_j^{(*)}\in C\ell_6, j=1, 2, 3 \, (\alpha$ and j playing the role of flavour (weak isospin) and colour index, respectively).

\smallskip

It was observed in \cite{DT20} that the projector $\ell + \mathfrak{q}$ on leptons and quarks,
\be 
\label{leptons+quarks}
\ell=p_1p_2p_3, \, \mathfrak{q}=\sum_{j=1}^3 q_j, \, q_j=p_jp_k^{'}p_\ell^{'} \, (j,k,\ell)\in Perm(1,2,3); \,p_j=b_jb_j^*, p_k^{'}=b_k^*b_k, 
\ee
kills $b_j^{(*)}$ and, more generally, the odd part $C\ell_6^1$ of $C\ell_6$: 
\be 
\label{KillOddb}
(\ell+\mathfrak{q})C\ell_{10}(\ell+\mathfrak{q})\subset C\ell_4\otimes C\ell_6^0 \, \, \, ((\ell+\mathfrak{q})b_j^{(*)}(\ell+\mathfrak{q})=0).
\ee

In the present paper we promote the exactly conserved weak hypercharge to a superselection rule: \textit{Y commutes with all observables and all symmetry transformations}, and explore its consequences. As a first corollary we obtain  a $u(1)$ extension of the gauge Lie algebra of the SM: the centralizer of Y in $so(10)$ is  $\mathfrak{g}=u(2)\oplus u(3)$ (\ref{u1u3}). The Lie algebra $\mathfrak{g}_{SM}$ (\ref{GSM}) of the gauge group of the SM is the maximal subalgebra of $\mathfrak{g}$ that annihilates the rank two Jordan subalgebra $J_{s\nu}$ of sterile (anti) neutrinos. The superselection rules forbid coherent superpositions of quantum states with different values of superselected charges, \cite{G}. Conversely, we only distinguish
particles from antiparticles if they carry different eigenvalues of Y. (Allowing the existence of a Majorana neutrino one cannot speak of a right chiral neutrino $\nu_R$ or of its antiparticle, both having $Y=0$.) We are thus led to project on a 15 dimensional particle subspace, excluding the right chiral neutrino $\nu_R=\ell \pi_1\pi_2, \, \pi_\alpha = a_\alpha 
a^*_\alpha$. This changes in an interesting way the Higgs field, identified with the scalar part of a superconnection, which belongs to $C\ell_4^1$, and the associated Lie superalgebra:
\be 
\label{a*transforme}
\mathcal{P}=\ell(1-\pi_1\pi_2)\Rightarrow  \mathcal{P}a_\alpha^{(*)} \mathcal{P} = \ell a_\alpha^{(*)} \pi_{\bar{\alpha}}^{'} + \mathfrak{q} a_\alpha^{(*)}, \, \bar{\alpha}=3-\alpha. \,
\ee
The leptonic part of the transformed Fermi oscillators (the term proportional to $\ell$) is identified with the odd generators of the simple Lie superalgebra $s\ell(2|1)$ applied over forty years ago by Ne'eman and Fairly to the Weinberg-Salam model. The fact that the lepton and the quark parts of the transformed $a_\alpha^{(*)}$ and of the Higgs field operator (\ref{Higgs}) differ, far from being a liability, yields (upon fixing the normalization $\rho$) the new relation
(\ref{cosWeinmerg}) between the Higgs and the W boson masses, in good agreement with their experimental values.  

\smallskip

The Yukawa coupling of fermions and the Higgs field, considered in Sect.~5 is a manifestation of {\it triality}: the coupling of the three $8$-dimensional representations of ${\rm Spin} (8)$ -- the left and right chiral spinors, corresponding to six (up and down colour) quarks and two leptons each, and an eight vector in internal space, to which we associate the Higgs superconnection. To display it one may use the octonion realization of ${\rm Spin} (10)$ acting on ${\mathbb C} \otimes {\mathbb O}^2$ described in \cite{Br}. In fact, the present approach was initiated in \cite{DV} by suggesting the Albert (or exceptional Jordan) algebra $J_3^8$ as a natural framework for displaying the three generations. It was soon realized that the Jordan subalgebra $J_2^8\subset J_3^8$ with automorphism group $Spin(9)$ \cite{TD, TDV} corresponds to one generation; then the  Clifford algebra $C\ell_9$ is privileged as an associative envelope of $J_2^8$ \cite{DT}. (The significance of $Spin(9)$ was further emphasized in \cite{K}; for its role in octonionic geometry - see also \cite{PP}.) 

\smallskip

Considering the complexification ${\mathbb C} \otimes J_3^8$ as a Jordan module, whose automorphism group is the compact $E_6 \, [Y]$, Boyle \cite{B} observes that ${\rm Spin} (10)$ naturally appears as its subgroup corresponding to one generation of fermions -- stabilizing an 1-dimensional projector in $J_3^8$. The particle ${\rm Spin} (10)$ module ${\mathbb C}^{16}$ is then realized as a subspace of the complex fundamental representation $\underline{27}$ of $E_6$.

\smallskip

The problem of incorporating in a meaningful way the three families of fundamental fermions into, say, a multiplicity three module of ${\mathbb C} \otimes J_3^8$ is still a challenge.

\bigskip

{\footnotesize\noindent {\bf Acknowledgments.} I have been induced  to think about the Jordan algebra approach to finite quantum geometry in numerous conversations with Michel Dubois-Violette, prior the publication of \cite{DV} and during our joint work with him afterwards. Stimulating discussions with Svetla Drenska, Kirill Krasnov and Jean Thierry-Mieg are also gratefully acknowledged. The author thanks IHES for hospitality during the final stage of this work when the chiral fermionic Lagrangian and the action of $Spin(10)$ on $\mathbb{C}\otimes \mathbb{O}^2$ were included.}

\bigskip

\appendix
\section{Superselected euclidean Jordan algebras}
\setcounter{equation}{0}
Euclidean Jordan algebras are commutative, power associative and partially ordered - by declaring the square of each element positive. (For a concise review and references - see Sect. 2 of \cite{T}.) Finite dimensional euclidean Jordan algebras can be decomposed into direct sums of simple ones, $J_r^d$, labelled by their rank r and degree d. Each $J_r^d$, viewed as a real vector space, splits into a direct sum of r 1-dimensional spaces $E_{ii}=\mathbb{R}e_i, e_i^2=e_i, i=1, ..., r,$ and $r\choose 2$ d-dimensional spaces $E_{ij}, 1\leq i<j\leq r, \, E_{ij}\circ E_{jk}\subset E_{ik}$. The $E_{ij}, i\leq j$ are eigenspaces of the left multiplication operators $L_{e_i} (L_xy=x\circ y=y\circ x)$ satisfying
\be 
\label{EigenEij}   
2L_{e_i} E_{jk} = (\delta_{ij} +\delta_{ik}) E_{jk} \Rightarrow e_i\circ e_j=\delta_{ij}e_j,\, E =\sum_{i=1}^r e_i =\textbf{1}_r
\ee
(i.e. $E(=E_{(r)})$ plays the role of the unit operator in the simple Jordan algebra $J_r^d$).
Such a splitting is called the \textit{Peirce decomposition} of $J_r^d$ (Benjamin Peirce, 1809-1880, was most of his life a Harvard professor). For $r\geq 2$ the range of d always includes the dimensions 1, 2, 4 of the associative division algebras. We are only dealing here with the complex number case, $d=2$, and with \textit{special Jordan algebras} - Jordan subalgebras of associative algebras in which the Jordan product is the symmetrized associative product: $2x\circ y =xy+yx$.

The simple Jordan subalgebras of $J_\mathcal{P}$ are labelled by the eigenvalues of Y. As the spectra of Y for left and right chiral fermions (with the sterile (anti)neutrino excluded) do not overlap, left and right observables belong to different simple components. It is convenient to order the irreducible components according to their rank. There is a single rank one Jordan subalgebra with $Y=-2$, corresponding to the right chiral electron: $J_1 = \mathbb{R} e_R, e_R=\pi_1^{'}\pi_2^{'}\ell$.
As there are two rank two algebras in our list, we shall first describe the general $J_2^2$ algebra and then specify  its different physical interpretations. The Weyl basis of the real 4-dimensional algebra $J_2^2=\mathcal{H}_2(\mathbb{C})$ can be written as $e_\alpha, \alpha=1, 2$ and $e_{12}= S_+, e_{21}=S_-=S_+^* (e_{12}$ and $e_{21}$ being two conjugate to each other
generators of the real 2-dimensional Peirce subspace $E_{12}$) satisfying
 $$
 e_1 S_+ = S_+ = S_+e_2, \, e_2S_+=0=S_+e_1, \, e_2S_-=S_-=S_-e_1, \, e_1S_-=0=S_-e_2, 
 $$
 \be 
 \label{J22S}
S_\pm^2=0, \, e_1=S_+S_-, \, e_2=S_-S_+, \, e_\alpha e_\beta = \delta_{\alpha \beta} \, e_\alpha.
\ee
The two realizations of $J_2^2$ in our list are $J_{s\nu}$ with $e_1=\nu_R, e_2=\bar{\nu}_L, \, S_+=\Omega$ and $J_2^L$ with
$e_1=\nu_L, e_2=e_L,\, S_\pm = I_\pm$. In any $J_2^2$ we have a family of hermitian elements $S_\vartheta\in E_{12}$ spanning the unit circle in $E_{12}$:
\be 
\label{Stheta}
S_\vartheta = e^{i\vartheta}S_+ + e^{-i\vartheta}S_-, -\pi<\vartheta\leq \pi; \, S_\vartheta^2=e_1+e_2= \textbf{1}_2 .
\ee
There are two simple Jordan components of type $J_3^2$ of the algebra of superselected observables in $C\ell_{10}$. They both correspond to right chiral quarks differing by their flavour projections. We shall give a unified description of $J_3^2$ in the 3-dimensional space of colour degrees of freedom. If $(j, k, \ell)$ is a permutation of $(1, 2, 3)$ then a natural Weyl basis in  $J_3^2$ reads (cf. (\ref{Pbj*k})):
\be 
\label{J32Weyl}
e_j=q_j=p_jp_k^{'}p_l^{'}, \, e_{jk}=B_{kj} = b_k^*b_jp_\ell^{'}\Rightarrow q_jB_{kj}q_k=B_{kj}, B_{kj}B_{jk}=q_j.
\ee
As noted in the introduction, it is no surprise that one finds here the same expressions (\ref{Pbj*k}) encountered in computing the generators of the (complexified) $su(3)$ Lie algebra. Finally, the Weyl basis for$J_6^2$ can be written in terms of tensor products of elements of $J_2^2$ and $J_3^2$. In particular, the primitive idempotents corresponding to left chiral coloured quarks (\ref{quarks}) are products of $e_1=\pi_1^{'}\pi_2$ and $e_2=\pi_1\pi_2^{'}$ with $q_j$; a typical off diagonal element reads:
$$
e_{1j,2k} = a_1^*a_2b_jb_k^*p_\ell^{'} = u_L^je_{1j, 2k}d_L^k.
$$ 
 


\newpage

\begin{thebibliography}{000}
\bibitem[Ab]{Ab} R. Ablamowicz, Construction of spinors via Witt decomposition and primitive idempotents: a review, \textit{Clifford algebras and spinor structures}. Kluwer Acad. Publ., 1995; -, On the structure theorem of Clifford algebras, arXiv:1610.02418 [math.RA].
\bibitem[B20]{B20} J.C. Baez, Getting to the bottom of Noether theorem, arXiv:2006.14741 [hep-th].
\bibitem[BH]{BH} J.C. Baez, J. Huerta, The algebra of grand unified theory, {\it  Bull. Amer. Math. Soc.} {\bf 47}:3 (2010) 483-552; arXiv:0904.1556v2 [hep-th].
\bibitem[B]{B} L. Boyle, The standard model, the exceptional Jordan algebra and triality, arXiv:2006.16265 [hep-th].   
\bibitem[BF]{BF} L. Boyle, S. Farnsworth, The standard model, the Pati-Salam model, and ``Jordan
geometry''. arXiv:1910.11888 [hep-th].
\bibitem[BS]{BS} A. Bochniak, A. Sitarz, A spectral geometry for the Standard Model without fermion doubling, arXiv:2001.02902 [hep-th].
\bibitem[Br]{Br} R.L. Bryant,Notes on spinors in low dimensions, arXiv:2011.05568.
\bibitem[CC]{CC} A.H. Chamseddine, A. Connes, Noncommutative geometry as a framework for unification of all fundamental interactions including gravity, {\it Fortschr. Phys.} {\bf 58} (2010) 553-600; arXiv:1004.0464 [hep-th].
\bibitem[D]{D} M. Dubois-Violette, \newblock Complex structures and the {E}lie {C}artan approach to the theory of
  spinors, \, in: {\em Spinors, Twistors, Clifford Algebras and   Quantum Deformations}, pp. 17--23, Kluwer Acad., 1993; hep-th/9210108.
\bibitem[DV]{DV} M. Dubois-Violette, Exceptional quantum geometry and particle physics, {\it Nucl. Phys.} B {\bf 912} (2016) 426-444; arXiv:1604.01247 [hep-th].
\bibitem[DT]{DT} M. Dubois-Violette, I. Todorov, Exceptional quantum geometry and particle physics II, {\it Nucl. Phys.} B {\bf 938} (2019) 751-761; arXiv:1808.08110.
\bibitem[DT20]{DT20} M. Dubois-Violette, I. Todorov, Superconnection in the spinfactor approach to particle physics, \textit{Nuclear Phys.} \textbf{B957} (2020) 115065; arXiv:2003.06591 [hep-th].
\bibitem[F19]{F19} K. Fredenhagen, Independent quantum systems and the associativity of the product of quantum observables, {\it Philosophical Problems in Science} (ZFN) {\bf 66} (2019) 61-72.
\bibitem[F]{F} C. Furey, $SU (3) c \times SU (2) L \times U (1) Y (\times U (1) X )$ as a symmetry of the division algebra ladder operators, {\it Eur. Phys. J.} {\bf C78} (2018) 375; arXiv:1806.00612 [hep-th].
\bibitem[G]{G} D. Giulini, Superselection rules, arXiv:0710.1516v2 [quant-ph].
\bibitem[GQS]{GQS} G. G\"otz, T. Quella, V. Schomerus, Representation theory of $s\ell(2|1)$, \textit{Jour. of Algebra} \textbf{312} (2007) 829-848; hep-th/0504234v2.
\bibitem[GIS]{GIS} J.M. Gracia-Bondia, B. Iochum, T. Schucker, The Standard Model in noncommutative geometry and fermion doubling, {\it Phys. Lett.} B {\bf 416} (1998) 123; hep-th/9709145.
\bibitem[H]{H} R. Haag, {\it Local Quantum Physics, Fields, Particles, Algebras}, Springer, Berlin, 1993.
\bibitem[H17]{H17} M.J.D. Hamilton, \textit{Mathematical Gauge Theory With Applications to the Standard Model of Particle Physics}, Springer Nature, 2017 (657 pages).
\bibitem[HLN]{HLN} D.S. Hwang, C.Y. Lee, Y. Ne'eman, BRST quantization of the $SU(2|1)$ electroweak theory in  the superconnection approach to the Higgs mass, \textit{Int. J. Mod. Phys.}  \textbf{A11} (1996) 3509-3522.  
\bibitem[JvNW]{JvNW} P.~Jordan, J.~von Neumann, and E.~Wigner.
\newblock On an algebraic generalization of the quantum mechanical formalism,
\newblock {\em Ann. Math.}, \textbf{36} (1934) 29--64.
\bibitem[K13]{K13} A. Kapustin, Is quantum mechanics exact? \textit{J. Math. Phys.} \textbf{54} (2013) 062107; -, Is there life beyond quantum mechanics? arXiv:1303.6917v4. 
\bibitem[KS]{KS} G.K. Karananas, M. Shaposhnikov, Gauge coupling unification without leptoquarks, \textit{Phys. Lett.} B \textbf{771} (2017) 332-338; arXiv:1703.02464v2.
\bibitem[K]{K} K. Krasnov, $SO(9)$ characterization of the standard model gauge group, arXiv:1912.11282.
\bibitem[M]{M} J. Maldacena, The symmetry and simplicity of the laws of physics and the Higgs boson, arXiv:1410.6753v2 [physics.pop-ph]
\bibitem[MQ]{MQ} V. Matthai, D. Quillen,  Superconnections, Thom classes, and covariant differential forms, {\it Topology} {\bf 25} (1986) 85-110.
\bibitem[PP]{PP} M. Parton, P. Piccinni, The role of Spin(9) in octonionic geometry, \textit{Axioms} \textbf{7}:4 (2018) 72; arXiv:1810.06288 [math.DG]. 
\bibitem[PDG]{PDG} Particle Data Group, CKM-quark mixing matrix, Revised March 2020 by A. Ceccucci, Z. Ligeti, Y. Sakai,
\textit{Prog. Theor. Exp. Phys.} \textbf{2020}, 083C01.
\bibitem[P]{P} B. Pontecorvo, Neutrino experiment and the problem of conservation of leptonic charge, \textit{Zh. Eksp. Teor. Fiz.}  \textbf{53} (1967) 1717-1725 (English transl.\textit{Sov. Phys. JETP} \textbf{26}:5 (1968) 984-988).  
\bibitem[Q]{Q} D. Quillen, Superconnections and the Chern character, {\it Topology} {\bf 24} (1985) 85-95.
\bibitem[R]{R} G. Roepstorff, Superconnections and the Higgs field, {\it J. Math. Phys.} {\bf 40} (1999) 2698-2715; arXiv:
hep-th/9801040.
\bibitem[S]{S} W.D. van Suijlekom, {\it Noncommutative Geometry and Particle Physics},
Springer, Dordrecht 2015.
\bibitem[T-MN]{T-MN} J. Thierry-Mieg, Y. Ne'eman, Exterior gauging of an internal symmetry and $SU(2M)$ quantum asthenodynamics,
\textit{Proc. Nat. Acad. Sci} \textbf{79} (1982) 7068-7072.
\bibitem[T-M]{T-M} J. Thierry-Mieg, Chirality, the missing key to the definition of the connection and curvature of a Lie-Kac superalgebra, \textit{JHEP} \textbf{01} (2021) 111; arXiv:2003.12234 [hep-th].
\bibitem[T-M20]{T-M20} J. Thierry-Mieg, Scalar anomaly cancellation reveals the hidden algebraic structure of the quantum chiral $SU(2|1)$ model of leptons and quarks, \textit{JHEP} \textbf{10} (2020) 167; arXiv:2005.04754v2 [hep-th].
\bibitem[T]{T} I. Todorov, Exceptional quantum algebra for the standard model of particle physics, in: \textit{Lie Theory and Its Applications in Physics}, V.K. Dobrev (Ed.), Chpt. 3, Springer Nature, 2020; 1911.13124 [hep-th]; \, -, Jordan algebra approach to finite quantum geometry, \textit{PoS} (CORFU 2019)163.
\bibitem[TD]{TD} I. Todorov, S. Drenska, Composition algebras, exceptional Jordan algebra and related groups, \textit{JGSP} \textbf{46} (2017) 59-93; \, - -, Octonions, exceptional Jordan algebra, and the role
of the group $F_4$ in particle physics, {\it Adv. in Appl. Clifford Alg.} {\bf 28} (2018) 82;
arXiv:1805.06739v2 [hep-th].
\bibitem[TDV]{TDV} I. Todorov, M. Dubois-Violette, Deducing the symmetry of the standard model from the automorphism and structure groups of the exceptional Jordan algebra, {\it Int. J. Mod. Phys.} A {\bf 33} (2018) 1850118; arXiv:1806.09450.
\bibitem[WWW]{WWW} G.C. Wick, A.S. Wightman, E.P. Wigner, The intrinsic parity of elementary particles, \textit{Phys. Rev.} \textbf{88}:1 (1952) 101-105; -, -, -, Superselection rule for charge, \textit{Phys. Rev.} \textbf{D1}:12 (1970) 3267-3269.
\end{thebibliography}
\end{document}